\documentclass[aps,pre,twocolumn]{revtex4-1}
\usepackage{graphicx}
\usepackage{dcolumn}
\usepackage{bm}
\usepackage{amsmath}
\usepackage{mathrsfs}
\begin{document}

\title{Fractal and Small-World Networks Formed by Self-Organized Critical Dynamics}

\author{Akitomo Watanabe}
\email{akitomo0416watanabe@eng.hokudai.ac.jp}
\author{Shogo Mizutaka}
\email{s.mizutaka@eng.hokudai.ac.jp}
\author{Kousuke Yakubo}
\email{yakubo@eng.hokudai.ac.jp} \affiliation{Department of Applied Physics, Graduate
School of Engineering, Hokkaido University, Sapporo 060-8628, Japan}

\date{\today}
\begin{abstract}
We propose a dynamical model in which a network structure
evolves in a self-organized critical (SOC) manner and explain a
possible origin of the emergence of fractal and small-world
networks. Our model combines a network growth and its decay by
failures of nodes. The decay mechanism reflects the instability
of large functional networks against cascading overload
failures. It is demonstrated that the dynamical system surely
exhibits SOC characteristics, such as power-law forms of the
avalanche size distribution, the cluster size distribution, and
the distribution of the time interval between intermittent
avalanches. During the network evolution, fractal networks are
spontaneously generated when networks experience critical
cascades of failures that lead to a percolation transition.
In contrast, networks far from criticality have small-world
structures. We also observe the crossover behavior from fractal
to small-world structure in the network evolution.
\end{abstract}

\maketitle

\section{Introduction}
\label{sec:intro}

Complex systems consisting of discrete elements and their pair
interactions can be described by networks. Many of large-scale
networks representing complexity of the real world are known to
have common properties in their topology
\cite{Albert02,Dorogovtsev03,Cohen10}, such as the
scale-free property \cite{Barabasi99}, degree correlations
\cite{Newman02,Newman03}, or community structures
\cite{Girvan02}. In particular, structures of real-world
networks are classified into two types from a viewpoint of the
relation between the number of nodes and the path length,
namely small-world structures \cite{Watts98} and fractal
structures \cite{Song05}. For a small-world network, the
average path length $\langle l\rangle$ is extremely small
comparing to the network size $N$ and increases at most
logarithmically with $N$, i.e., $\langle l \rangle \propto \log
N$. Numerous real-world complex networks possess the
small-world property \cite{Montoya02,Amaral00,Bassett06,
Humphries08}. On the other hand, a network is called fractal if
the relation $N_{\text{B}}(l_{\text{B}})\propto
l_{\text{B}}^{-d_{\text{B}}}$ holds, where
$N_{\text{B}}(l_{\text{B}})$ is the minimum number of subgraphs
of diameter less than $l_{\text{B}}$ required to cover the network
and $d_{\text{B}}$ is the fractal dimension \cite{Song05}. Since
this relation at $l_{\text{B}}\sim \langle l \rangle$ suggests
the power-law scaling $\langle l \rangle \propto
N^{1/d_{\text{B}}}$ \cite{Kawasaki10}, the fractal nature seems
to conflict with the small-world property. Nevertheless, real
complex networks that are small world in the sense of $\langle
l \rangle \propto \log N$ often satisfy the fractal scaling
$N_{\text{B}}(l_{\text{B}})\propto
l_{\text{B}}^{-d_{\text{B}}}$, as observed in the world-wide
web, actor networks, protein interaction networks, cellular
networks \cite{Song05}, power-grid networks \cite{Csanyi04},
and software networks \cite{Concas06,Myers03}. This apparent
inconsistency can be reconciled by taking into account a
structural crossover from fractal to small-world scaling
associated with the change in length scale \cite{Kawasaki10}.

It is well understood that the small-world property arises from
the existence of short-cut edges \cite{Watts98}. Only a tiny
amount of short-cut edges added into a non-small-world network
drastically reduces the average path length. In contrast, the
microscopic mechanism of the emergence of fractality in complex
networks still remains unclear though fractal networks and
their relation to the scale-free property have been extensively
studied \cite{Song05,Kawasaki10,Cohen04,Goh06,Song06,Rozenfeld07,
Song07,Furuya11,Sun14}. It is thus also not understood why there
exist small-world and fractal networks in the real world and
how fractal networks crossover to small-world ones. In order to
deal with these problems, it is significant to remind that many
conventional fractal objects embedded in the Euclidean space
are formed by dynamics exhibiting self-organized criticality
\cite{Bak87,Bak93,Markovic14,Drossel92,Takayasu92,Rinaldo93,
Sapoval04}. In self-organized critical (SOC) dynamics, a system
approaches spontaneously a critical point without tuning
external parameters and fluctuates around the critical state
due to the instability of the critical or near-critical states.
One of the remarkable features of SOC dynamics is that
stationary fluctuations around criticality are accompanied by
intermittent, avalanche-like bursts of some sort of dynamical
quantities, in which the avalanche size distribution obeys a
power law. It is natural to consider that fractal complex
networks are also formed by SOC dynamics.

SOC dynamics \textit{on} static complex networks have been
extensively studied in previous works \cite{Markovic14,deArcangelis02,
Moreno02,Goh03,Masuda05,Lin05,Pellegrini07, Luque08,Bhaumik13,
Watanabe14,Noel14}. However, for the construction of fractal
networks through SOC dynamics, we need to consider the
interplay between internal dynamics and the network topology
\cite{Gross08,Aoki12,Shimada14}, by which the network evolution
itself displays SOC characteristics. There have been many models
of network evolution driven by internal dynamics related to
self-organized criticality, such as models based on the
Bak-Sneppen dynamics \cite{Christensen98,Slanina99,Garlaschelli07},
other models of ecological systems \cite{Guill08}, models related
to the sandpile dynamics \cite{Peixoto04,Peixoto08,Fronczak06},
a model describing the motion of solar flares \cite{Hughes03},
and rewiring models based on state changes of nodes or edges
\cite{Bornholdt00,Rybarsch14,Bianconi04,Shin06}. Although these
models generate nontrivial networks through the couplings to
internal dynamics, it is difficult to say that fractal networks
are formed by SOC dynamics in these models because of the lack of
intermittent, avalanche-like behavior of network structures
\cite{Christensen98,Bornholdt00,Rybarsch14}, the necessity of 
parameter tuning for criticality \cite{Bianconi04,Garlaschelli07},
or the absence of fractality in generated networks \cite{Slanina99,
Hughes03,Bianconi04,Peixoto04,Peixoto08,Fronczak06,Shin06,
Guill08}.

In this paper, we present a model of fractal networks formed by
SOC dynamics. Taking into account the evolution of real
networks, the network instability required for SOC dynamics
is realized by overload failures of nodes. In general, when the
size of a functional network becomes large, the probability
that all nodes in the network can escape failures decreases.
Failure(s) on a single or a few nodes can cause a cascade of
overload failures, and the network decays into smaller ones.
Our model combines a network growth by introducing new nodes
and its decay due to the instability of large grown networks
against cascading overload failures. It is numerically demonstrated
that the present dynamical system exhibits self-organized
criticality and the network evolution generates both fractal
and small-world networks. Furthermore, the crossover behavior
from fractal to small-world structure has been observed in SOC
dynamics.

The rest of this paper is organized as follows. In Sect.~2, we
formulate the model combining a network growth with cascading
overload failures induced by fluctuating loads. Our numerical
results are presented in Sect.~3. In this section, we show the
time development of several measures describing the network
structure, SOC character of the dynamics, and the fractal and
small-world properties of networks generated in SOC dynamics.
Section 4 is devoted to the summary.

\section{Model}
\label{sec:model}

\subsection{Network instability --- Cascading overload failures}
\label{subsec:instability}

In the present work, the instability required to construct an
SOC model is realized by \textit{cascading overload failures}
in large networks. Our daily life is supported by various
functional networks, such as power grids, computer networks,
the world-wide web, etc. Functions of networks are achieved by
some sort of \textit{flow} which plays, at the same time, a
role of \textit{loads} in the network. The load on a node
usually fluctuates temporally and its instantaneous value
exceeding the allowable range causes a failure of node. This
overload failure may induce a cascade of subsequent failures
which reduces, sometimes greatly, the network size. Such
cascades of failures provide the instability of networks.
Recently, the robustness of a network against cascading
overload failures induced by temporally fluctuating loads has
been studied \cite{Mizutaka15}, This study employs the random
walker model proposed by Kishore et al.\cite{Kishore11,
Kishore12}, in which fluctuating loads are described by random
walkers moving on a network. Since our model is based on the
study by Ref.~\ref{mizu}, we briefly review this work.

In the random walker model \cite{Kishore11,Kishore12,Mizutaka13},
we consider $W_{0}$ non-interacting random walkers moving on a
connected and undirected network with $M_{0}$ edges. Walkers on
a node represent the temporally fluctuating load imposed on the
node. Since the stationary probability $p_{k}$ to find a walker
on a node of degree $k$ is given by $p_{k}=k/2M_{0}$
\cite{Noh04}, the probability that there exist $w$ walkers on
the degree-$k$ node is presented by
\begin{equation}
h_{k}(w) = \binom{W_{0}}{w} p_{k}^{w} {(1-p_{k})}^{W_{0} -w}.
\label{hkW0w}
\end{equation}
This binomial distribution gives the average load on the
degree-$k$ node as $\langle w\rangle_{k}=W_{0}p_{k}$ and the
standard deviation $\sigma_{k}=\sqrt{\langle w\rangle_{k}(1-p_{k})}$.
It is then natural to define the capacity of a node of degree
$k$ as
\begin{equation}
q_{k}= \langle w\rangle_{k} + m\sigma_{k},
\label{qkW0}
\end{equation}
where $m$ is a real positive parameter and characterizes the
tolerance of the node to load. A node is considered to fail if
the number of walkers $w$ on the node exceeds this capacity.
Therefore, the probability $F_{W_{0}}(k)$ that a node of degree
$k$ experiences an overload failure is calculated by summing up
the distribution function $h_{k}(w)$ over $w$ larger than $q_{k}$.
Using the regularized incomplete beta function $I_{x}(a,b)$ for
this summation \cite{Abramowitz64}, the overload probability is
expressed as \cite{Kishore11}
\begin{equation}
F_{W_{0}}(k) =I_{k/2M_{0}}(\lfloor q_{k}\rfloor +1,W_{0}-\lfloor q_{k}\rfloor ),
\label{eq:FWk}
\end{equation}
where the floor function $\lfloor x\rfloor$ represents the
greatest integer less than or equal to $x$.

Applying the above idea of the overload probability, a cascade
of failures starting with a specific large network can be described
as follows \cite{Mizutaka15}.
\begin{itemize}
\item[(i)] Prepare an initial connected, undirected, and
    uncorrelated network $\mathcal{G}_{0}$ with $N_{0}$
    nodes and $M_{0}$ edges, in which $W_{0}$ random
    walkers exist, where $W_{0}$ is chosen so as to be
    proportional to $M_{0}$. In addition, determine the
    capacity $q_{k}$ of each node by Eq.~(\ref{qkW0}).
\item[(ii)] At each time step $\tau$, assign $W_{\tau}$
    random walkers to the network $\mathcal{G}_{\tau}$ at
    step $\tau$, where the total load $W_{\tau}$ is given
    by
    \begin{equation}
    W_{\tau}=W_{0}\left( \frac{M_{\tau}}{M_{0}} \right)^{r}.
    \label{eq:Wtau}
    \end{equation}
    Here $M_{\tau}$ is the total number of edges in the
    network $\mathcal{G}_{\tau}$ and $r$ is a real positive
    parameter.
\item[(iii)] Calculate the overload probability of every
    node, and remove nodes from $\mathcal{G}_{\tau}$ with
    this probability.
\item[(iv)] Repeat (ii) and (iii) until no node is removed
    in the procedure (iii).
\end{itemize}
The reduction of the total load in the procedure (ii)
corresponds to actual cascades of failures during which the
total load is reduced to some extent to prevent the breakdown
of the network function. We call the exponent $r$ in
Eq.~(\ref{eq:Wtau}) the load reduction parameter hereafter. It
should be emphasized that the overload probability in the
procedure (iii) cannot be calculated by Eq.~(\ref{eq:FWk}) with
$W_{0}$ replaced by $W_{\tau}$ for the following two reasons.
First, the degree $k$ of a node in the network
$\mathcal{G}_{\tau}$ is not the same with its initial degree
$k_{0}$. The capacity of a node is determined by the initial
degree $k_{0}$, while the probability to find a walker on this
node is proportional to the present degree $k$. Thus, the
overload probability depends on both $k$ and $k_{0}$. Secondly,
the network $\mathcal{G}_{\tau}$ is not necessarily connected,
though the initial network $\mathcal{G}_{0}$ is connected. If
$\mathcal{G}_{\tau}$ is not connected, $W_{\tau}$ random
walkers are distributed to each component in proportion to the
number of edges in the component, before starting the next
cascade step. Taking into account these remarks and the fact that
random walkers cannot jump to other components, the overload
probability of a node of degree $k$, whose initial degree is
$k_{0}$, in the $\alpha$-th component of $\mathcal{G}_{\tau}$
is given by
\begin{equation}
F_{W_{\tau}^{\alpha}}(k_{0},k)=I_{k/2M_{\tau}^{\alpha}}
\left(\lfloor q_{k_{0}}\rfloor +1, W_{\tau}^{\alpha}-\lfloor q_{k_{0}}\rfloor \right),
\label{eq:Ftaualpha}
\end{equation}
where $M_{\tau}^{\alpha}$ is the number of edges in the
$\alpha$-th component of $\mathcal{G}_{\tau}$ and
$W_{\tau}^{\alpha}=W_{\tau}M_{\tau}^{\alpha}/M_{\tau}$ is the
load assigned to the $\alpha$-th component. Since $k$ is always
equal to $k_{0}$ and the network $\mathcal{G}_{\tau}$ is
connected at $\tau=0$, $F_{W_{\tau}^{\alpha}}(k_{0},k)$ at
$\tau=0$ coincides with $F_{W_{0}}(k)$ given by
Eq.~(\ref{eq:FWk}). Thus, Eq.~(\ref{eq:Ftaualpha}) is a general
form of the overload probability for $\tau\ge 0$.

The relative size $S_{\text{f}}$ of the giant component in the
network $\mathcal{G}_{\text{f}}$ at the final stage of the
cascade process is an important quantity to evaluate the
robustness of networks against cascading overload failures.
This quantity $S_{\text{f}}$ can be calculated by combining the
generating function method \cite{Newman01} and the master
equation for the probability $\Pi_{\tau}(k_{0},k)$ that a node
in $\mathcal{G}_{\tau}$ has the present degree $k$ and the
initial degree $k_{0}$, without simulating numerically the
cascade process (i)--(iv) \cite{Mizutaka15}. By means of this
method, it has been clarified that there exists a threshold
value of the load reduction parameter $r_{\text{c}}(N_{0})$
above which $S_{\text{f}}$ becomes finite and below which
$S_{\text{f}}=0$ and $r_{\text{c}}(N_{0})$ for $N_{0}\to
\infty$ provides a percolation transition by cascading overload
failures \cite{Mizutaka15}. The critical property of
$\mathcal{G}_{\text{f}}$ at $r=r_{\text{c}}$ has also been
confirmed by the fractality of the giant component in
$\mathcal{G}_{\text{f}}$. These facts will be closely related
to the results of the present work. However, in the case that
the structure and node capacities of a network with which the
cascade starts depend on results of cascading failures
occurring in the past, as in the model of this work, it is
unfortunately impossible to apply the method utilizing the
generating function. In such a case, we need to simulate
numerically the process (i)--(iv) to find the final network
state after a cascade.

\subsection{Network evolution}
\label{subsec:evolution}

The basic idea of our dynamical model is to combine the growth
of a network and its decay into smaller ones by cascading
overload failures. The overload probability $F_{W_{0}}(k)$
given by Eq.~(\ref{eq:FWk}) is almost independent of $M_{0}$ if
$W_{0}\propto M_{0}$. However, the probability that the first
failures inducing a cascade of subsequent failures occur in the
network increases with the network growth. Thus, we expect that
the network cannot grow infinitely and its size fluctuates
around a certain value. Since our model includes two different
types of dynamics, i.e., growth and cascade of failures, in
order to distinguish them clearly, a time step of the network
growth is hereafter denoted by $t$ in parentheses and a cascade
step by subscript $\tau$. The concrete algorithm of the network
evolution in our model is then given as follows:
\begin{itemize}
\item[(1)] Start with a small and connected network
    $\mathcal{G}(0)$ with $N_{\text{ini}}$ nodes and
    $M_{\text{ini}}$ edges, in which $W_{\text{ini}}$
    random walkers exist. $W_{\text{ini}}$ is set as
    $W_{\text{ini}}=a M_{\text{ini}}$, where $a$ is a
    positive constant. The capacities of the nodes in
    $\mathcal{G}(0)$ are calculated by Eq.~(\ref{qkW0}).
\item[(2)] At every time step $t\ge 1$, add a new node with
    $\mu$ edges, where $\mu$ is in the range of $2\le \mu
    \le N_{\text{ini}}$, and connect the new node to $\mu$
    different nodes selected randomly from the network
    $\mathcal{G}(t-1)$ at time $t-1$. Let
    $\mathcal{G}_{0}(t)$ be the network at this stage.
\item[(3)] Place $W_{0}(t)=aM_{0}(t)$ random walkers on the
    network $\mathcal{G}_{0}(t)$, where $M_{0}(t)$ is the
    number of edges in $\mathcal{G}_{0}(t)$, and calculate
    the capacity of the new node by using Eq.~(\ref{qkW0})
    with $k$ replaced by $\mu$ and for the total load
    $W_{0}(t)$.
\item[(4)] Perform cascading overload failures starting
    with $\mathcal{G}_{0}(t)$ in accordance with the
    process (i)--(iv) described in
    Sect.~\ref{subsec:instability}. In this cascade
    process, isolated zero-degree nodes generated by the
    elimination of all their adjacent nodes, in addition
    to the overloaded nodes themselves, are removed from
    the system. Let $\mathcal{G}(t)$ be the resultant
    network after completing the cascade.
\item[(5)] Repeat the procedure from (2) to (4) for a
    sufficiently long period.
\end{itemize}

We should make several remarks concerning the above algorithm.
In the procedure (2), the number of edges $\mu$ of a newly
added node must be larger than $2$. Otherwise, once the network
is divided into disconnected components by cascading failures,
components never merge. The network after a long time becomes
an assembly of a large number of small graphs. Thus, the
dynamical system has a qualitatively different property from
that for $\mu\ge 2$. It should be also emphasized that the
network $\mathcal{G}_{0}(t)$ is not necessarily connected. If
$\mathcal{G}_{0}(t)$ consists of plural components, the total
load $W_{0}(t)$ is distributed to each component in the same
way as the case of cascading failures. Namely, the number of
walkers in the $\alpha$-th component is allocated by
\begin{equation}
W_{0}^{\alpha}(t)=W_{0}(t)\left[\frac{M_{0}^{\alpha}(t)}{M_{0}(t)}\right]
              =aM_{0}^{\alpha}(t),
\label{Wtalpha}
\end{equation}
where $M_{0}^{\alpha}(t)$ is the number of edges in the
$\alpha$-th component of $\mathcal{G}_{0}(t)$. The calculation
of the new-node capacity in the procedure (3) is actually done
by using $W_{0}^{\alpha}(t)$, because random walkers cannot
move to other components. The capacity of the node that is
introduced at time $t$ and embedded in the $\alpha$-th
component is then presented by
\begin{equation}
q^{\alpha}(t)= \langle w \rangle_{\mu}+m{\sigma}_{\mu}^{\alpha},
\label{q_new}
\end{equation}
where $\langle w\rangle_{\mu}=W_{0}^{\alpha}(t)p_{\mu}^{\alpha}=a\mu/2$,
$\sigma_{\mu}^{\alpha}=\sqrt{\langle w\rangle_{\mu}(1-p_{\mu}^{\alpha})}$,
and $p_{\mu}^{\alpha}=\mu/2M_{0}^{\alpha}(t)$. Once the
node capacity is determined, the value of $q$ never changes
until the node is eliminated.

The process of cascading overload failures in the procedure (4)
basically follows the steps (i)--(iv) in
Sect.~\ref{subsec:instability} with replacing
$\mathcal{G}_{0}$, $N_{0}$, $M_{0}$, and $W_{0}$ by
$\mathcal{G}_{0}(t)$, $N_{0}(t)$, $M_{0}(t)$, and $W_{0}(t)$,
respectively, where $N_{0}(t)$ is the number of nodes in
$\mathcal{G}_{0}(t)$. In addition to the possibility of
$\mathcal{G}_{0}(t)$ being disconnected, there are two other
differences in the detailed treatment of the cascade process.
Firstly, the overload probability of a node cannot be written
as Eq.~(\ref{eq:Ftaualpha}). This is because the node capacity
given by Eq.~(\ref{q_new}) depends on when the node was
introduced in the system. Thus, the overload probability of the
node $i$ at cascade step $\tau$ is written as
\begin{equation}
F_{W_{\tau}^{\alpha}(t)}(i)=I_{k_{i}/2M_{\tau}^{\alpha}(t)}
\left(\lfloor q^{\beta}(t_{i})\rfloor +1, W_{\tau}^{\alpha}(t)-
\lfloor q^{\beta}(t_{i})\rfloor \right),
\label{eq:Ftaualpha_i}
\end{equation}
where $k_{i}$ is the degree of the node $i$, $t_{i}$ is the
time at which the node $i$ was introduced, $\alpha$ and $\beta$
are the indices of the components to which the node $i$ belongs
at the present cascade step $\tau$ and at $t_{i}$,
respectively, and $q^{\beta}(t_{i})$ is presented by
Eq.~(\ref{q_new}). The symbols $W_{\tau}^{\alpha}(t)$ and
$M_{\tau}^{\alpha}(t)$ represent the number of random walkers
and number of edges in the $\alpha$-th component of
$\mathcal{G}_{\tau}(t)$, where $\mathcal{G}_{\tau}(t)$ is the
network at the step $\tau$ of the cascade starting with
$\mathcal{G}_{0}(t)$. The second difference is in the load
reduction scheme during the cascade. In
Sect.~\ref{subsec:instability}, the total number of random
walkers is reduced in accordance with the reduction of the
network size during the cascade, as expressed by
Eq.~(\ref{eq:Wtau}), to prevent the breakdown of the network
function. It is actually difficult to reduce quickly the total
load when the network size becomes large. Taking into account
such realistic situations of cascading failures, the load
reduction parameter $r$ characterizing how quickly the total
load is reduced with the reduction of the network size should
decrease with $N_{0}(t)$. Therefore, the total load during the
cascade is reduced according to
\begin{equation}
W_{\tau}(t)=W_{0}(t)\left[\frac{M_{\tau}(t)}{M_{0}(t)}\right]^{r[N_{0}(t)]},
\label{Wtau_t}
\end{equation}
where $M_{\tau}(t)$ is the number of edges in
$\mathcal{G}_{\tau}(t)$ and $r[N_{0}(t)]$ is a decreasing
function of $N_{0}(t)$. Since large-scale cascades are more
likely to occur if $r$ is small, the property that
$r[N_{0}(t)]$ decreases with $N_{0}(t)$ prevents the network
from growing too large.

If all nodes are eliminated from the system in the procedure
(4), the network $\mathcal{G}(t)$ is reset to $\mathcal{G}(0)$
and continue the network evolution from the procedure (2). In
the present model, the time scale of cascades measured by the
step $\tau$ is assumed to be much faster than that of the
network growth measured by the step $t$, and the relaxation
time of random walkers in a network is further shorter than a
single cascade step. We concentrate, in this work, on the
temporal evolution of networks in the time scale of the network
growth. Therefore, information on the network $\mathcal{G}(t)$
at every end of the procedure (4) is recorded to investigate
the model.

\section{Results and Discussion}
\label{sec:results}

Our model includes several parameters and conditions. These are
the numbers of nodes $N_{\text{ini}}$ and edges
$M_{\text{ini}}$ in the initial network $\mathcal{G}(0)$, the
topology of $\mathcal{G}(0)$, the load carried by a single edge
$a$, the node tolerance parameter $m$, the number of edges of a
newly added node $\mu$, and the functional form of the load
reduction parameter $r(N)$. In this section, we fix these
parameters as follows: The initial network $\mathcal{G}(0)$ is
a triangular ring with $N_{\text{ini}}=3$ and
$M_{\text{ini}}=3$. The parameters $a$ and $\mu$ are set as
$a=2.0$ and $\mu=2$, respectively. The value of $m$ is chosen
from the range of $5.0\le m \le 7.0$. The function $r(N)$ is
set as
\begin{equation}
r(N)=
\begin{cases}
r_{\text{max}} & \text{for } 2\le N <N_{\text{ini}},\\
\displaystyle
\frac{r_{\text{max}}(N_{\text{max}}-N)}{N_{\text{max}}-N_{\text{ini}}}
        & \text{for } N_{\text{ini}}\le N < N_{\text{max}},\\
0       & \text{for } N\ge N_{\text{max}}.
\end{cases}
\label{r(N)}
\end{equation}
This function decreases from its maximum value $r_{\text{max}}$
to zero as $N$ increases. Since a cascade of overload failures
with $r=0$ eliminates all nodes in any network, $N_{\text{max}}$
gives a rough estimation of the maximum network size in the
dynamics. Here we set $N_{\text{max}}=1,000$ and
$r_{\text{max}}=1.0$. We will explain later the reason why
we adopt the above parameter values and discuss suitable ranges
for the model parameters to obtain SOC character in the network
evolution.

\begin{figure}[ttt]
\begin{center}
\includegraphics[width=0.48\textwidth]{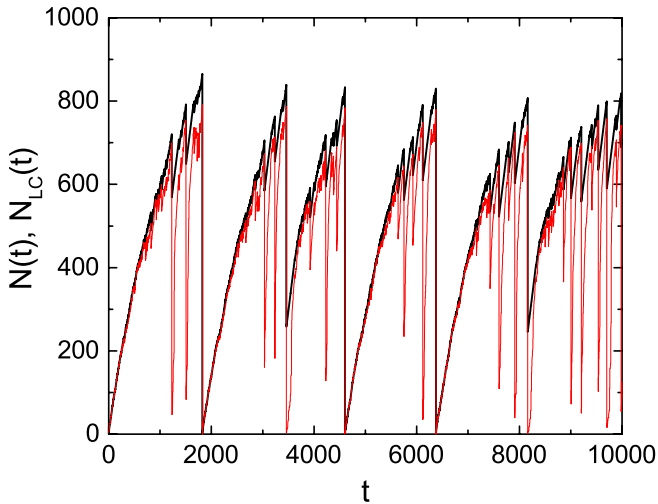}
\end{center}
\caption{(Color online) Time dependences of the number of nodes $N(t)$
in the network $\mathcal{G}(t)$ (thick black line) and number of nodes
$N_{\text{LC}}(t)$ in its largest component (thin red line). The node
tolerance parameter is set as $m=5.0$.}
\label{fig1}
\end{figure}
\subsection{Number of nodes and other network measures}
\label{subsec:network_size}

We first examine the time dependence of the size of the network
$\mathcal{G}(t)$. Figure 1 shows the number of nodes $N(t)$ in
$\mathcal{G}(t)$ for the first $10^{4}$ time steps. The result
clearly demonstrates that the network cannot grow infinitely
and the size $N(t)$ largely fluctuates by repetitive growth and
decay of the network. We can find some features in the line shape
of $N(t)$. In the early stage, $N(t)$ increases almost
monotonically with time, because the probability that any of
the nodes in the network fail is low due to small
$N_{0}(t)\, [=N(t-1)+1]$. Namely, in this time region, the
expectation number of failed nodes is less than $1$. After this
region, the expectation value becomes larger than $1$ and some
nodes fail in $\mathcal{G}_{0}(t)$. However, since
$r[N_{0}(t)]$ for still small $N_{0}(t)$ is rather large,
cascades are not widely spread. Thus, $N(t)$ for $400 \lesssim
t \lesssim 800$ keeps increasing with relatively small drops.
When $N(t)$ becomes larger than $800$, $r[N_{0}(t)]$ is so
small that a cascade of failures never stops until all nodes
are eliminated. Such \textit{complete collapses} occur at
$t=1,824$, $4,603$, and $6,374$ in Fig.~\ref{fig1}. After a
complete collapse, the system evolves in a similar manner to
the evolution from $t=0$. In addition to $N(t)$, we plot in
Fig.~\ref{fig1} the size of the largest component
$N_{\text{LC}}(t)$ contained in $\mathcal{G}(t)$. The size
$N_{\text{LC}}(t)$ basically follows the variation of $N(t)$ at
most of the time steps, but sometimes $N_{\text{LC}}(t)$ drops
substantially though $N(t)$ does not change so much. At these
times, the network is decomposed into small components by
cascading overload failures. The statistics of magnitudes of
drops in $N(t)$, i.e., cascade sizes, will be argued in the
next subsection.

\begin{figure}[ttt]
\begin{center}
\includegraphics[width=0.48\textwidth]{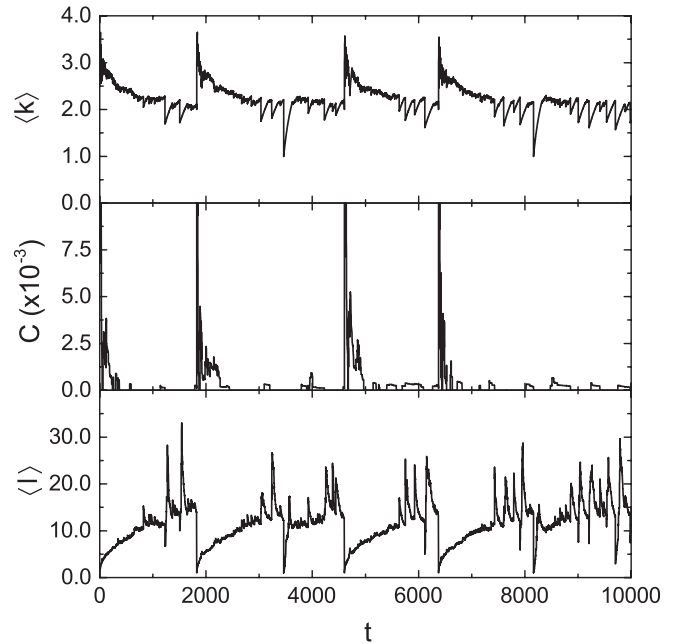}
\end{center}
\caption{Time dependences of the average degree (top), the
clustering coefficient (middle), and the average path length
(bottom) of the network $\mathcal{G}(t)$. The node tolerance
parameter is set as $m=5.0$. These quantities for the largest
component of $\mathcal{G}(t)$ are not shown in this figure,
because their line shapes almost overlap with those for
$\mathcal{G}(t)$.}
\label{fig2}
\end{figure}
We also calculated several quantities that characterize the
network structure at time $t$. Figure \ref{fig2} shows the
average degree $\langle k\rangle$, the clustering coefficient
$C$, and the average path length $\langle l\rangle$ of the
network $\mathcal{G}(t)$ as a function of $t$. These quantities
for the largest component of $\mathcal{G}(t)$ take almost the
same values as those for $\mathcal{G}(t)$. The average degree
fluctuates around $\langle k\rangle =2(=\mu)$ though it becomes
significantly larger than this value immediately after a
complete collapse. We have confirmed that the degree
distribution $\mathcal{P}(k)$ of $\mathcal{G}(t)$ hardly
depends on time and decays exponentially for large $k$ if
$N(t)$ is large enough (not shown here). This is reasonable
because random attachment of new nodes and cascading failures
without introducing degree correlations make the network
topology similar to a homogeneous random graph. The clustering
coefficient $C$ is quite small ($C\lesssim 10^{-4}$), except
for $\mathcal{G}(t)$ at and just after complete collapses. At a
complete collapse, $C$ is equal to $1$ because $\mathcal{G}(t)$
is a triangular ring at this time. The clustering coefficient,
however, rapidly decreases with the network growth by random
attachments. Sometimes $C$ becomes equal to zero, which implies
that the network takes a tree (or forest) structure. The
average path length $\langle l\rangle$ of the network at a
complete collapse is obviously $1$, and after that $\langle
l\rangle$ increases gradually with relatively large
fluctuations. Considering that $N(t)$ is less than $1,000$,
$\langle l\rangle$ close to or more than $10$ is too large to
regard the network as being small world. Then, we can expect
that $\mathcal{G}(t)$ giving very large $\langle l\rangle$ has
a fractal structure. Before discussing the fractality of
generated networks, it will be examined in the next subsection
whether our dynamical system exhibits SOC behavior.

\subsection{Avalanche size and self-organized criticality}
\label{subsec:SOC}

\begin{figure}[ttt]
\begin{center}
\includegraphics[width=0.48\textwidth]{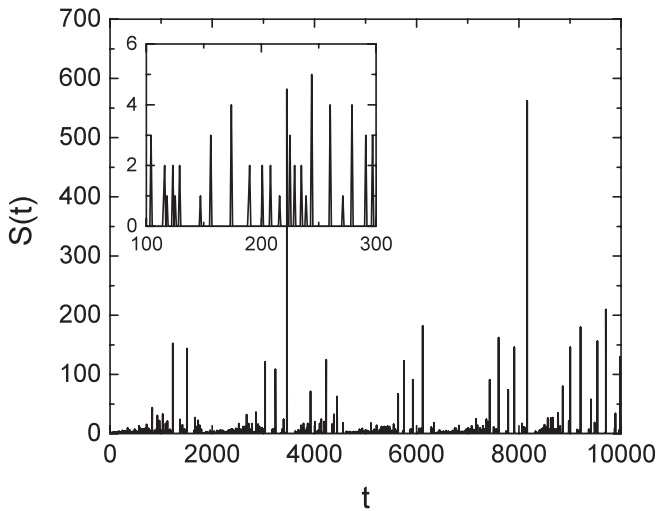}
\end{center}
\caption{Time dependence of the avalanche size $S(t)$ obtained
from $N(t)$ shown in Fig.~\ref{fig1}. To make the figure easy to
see, cascades leading to complete collapses are eliminated from
the figure. The intermittency of cascades is found in the inset
that magnifies the main figure for $100\le t \le 300$.}
\label{fig3}
\end{figure}
\begin{figure}[ttt]
\begin{center}
\includegraphics[width=0.48\textwidth]{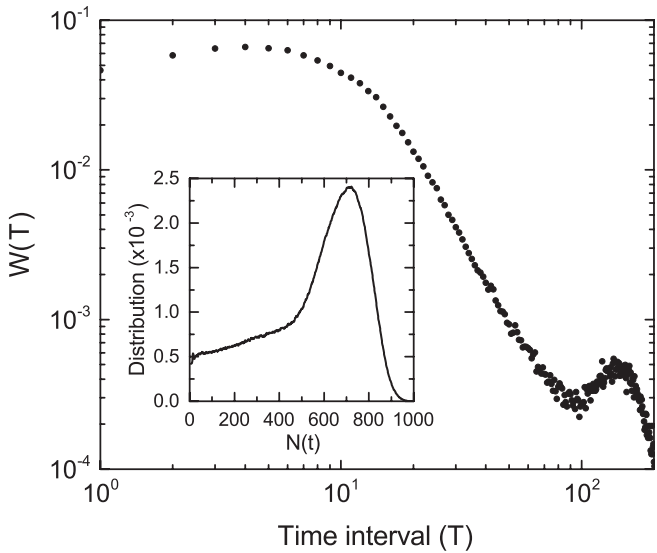}
\end{center}
\caption{Distribution function $W(T)$ of the inactive time
interval $T$ during which no overload failure occurs. The
distribution $W(T)$ is obtained from the dynamics up to
$t=5\times 10^{6}$ under the condition $m=5.0$. The inset
shows the distribution of $N(t)$ in the same dynamics.}
\label{fig4}
\end{figure}
Sudden drops of the network size $N(t)$ found in
Fig.~\ref{fig1} corresponds to decays of the network by
cascading overload failures. Magnitudes of these drops
represent scales of cascading overload failures. Here we
define the avalanche size $S(t)$ as the number of nodes that
are removed during a single cascade of overload failures
occurring at the time $t$. Figure \ref{fig3} shows the
avalanche size $S(t)$ obtained from $N(t)$ shown in
Fig.~\ref{fig1}. The avalanche size largely fluctuates even if
one ignores huge $S(t)$'s at complete collapses. Values of
$S(t)$ at most of the time steps are less than $50$, while
on rare occasions $S(t)$ exceeds $300$. The inset of
Fig.~\ref{fig3} demonstrates that these avalanches occur
intermittently with inactive intervals. This intermittency
suggests a possibility that the network dynamics possesses SOC
characteristics. In order to find further evidences of SOC
dynamics, we examine the distribution function $W(T)$ of the
inactive time interval $T$ between avalanches. The distribution
$W(T)$ obtained from the dynamics under the same conditions as
those for Fig.~\ref{fig1} but continued up to $5\times 10^{6}$
time steps is presented in Fig.~\ref{fig4}. This figure clearly
shows that $W(T)$ obeys a power law,
\begin{equation}
W(T)\propto T^{-\eta},
\label{eq:W(T)}
\end{equation}
in an intermediate region of $T$. The least-squares fit for the
data within $20\le T\le 50$ gives $\eta=3.00\pm 0.03$. The
small hump near $T\sim 150$ comes from a finite-size effect
related to the existence of the most probable network size
$N_{\text{typ}}$. This size is about $700$ for our choice of
the model parameters as depicted in the inset of Fig.~\ref{fig4}.
We have confirmed the correlations between $N(t)$ and $S(t)$
and between the inactive interval $T$ after a cascade and
the cascade (avalanche) size $S$. These correlations and
$N_{\text{typ}}\sim 700$ lead the frequently-appearing time
interval at $T\sim 150$.

\begin{figure}[ttt]
\begin{center}
\includegraphics[width=0.48\textwidth]{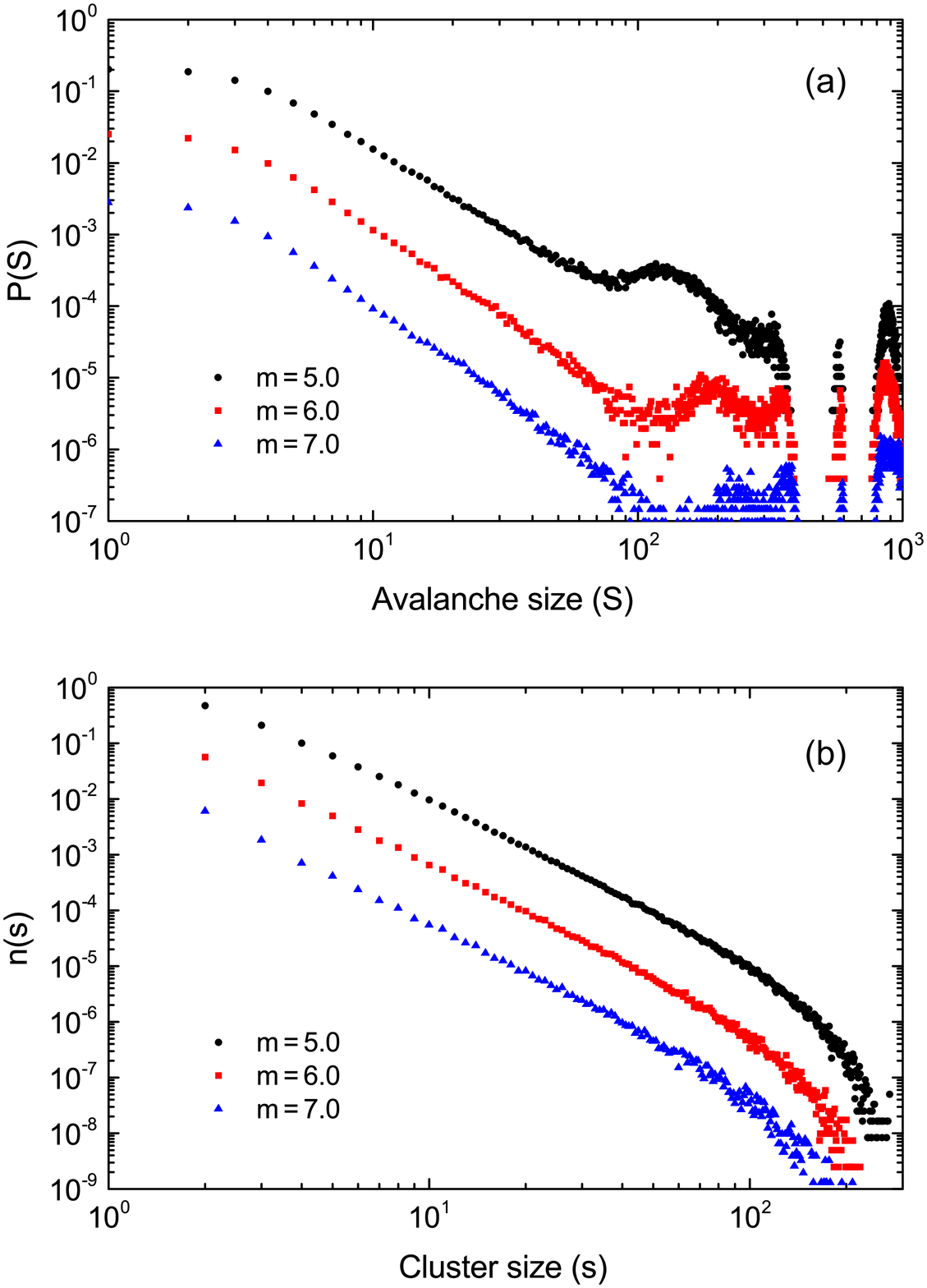}
\end{center}
\caption{(Color online) (a) Distribution function $P(S)$ of the
avalanche size $S$ and (b) the distribution function $n(s)$
of the cluster size $s$ for different values of the node
tolerance parameter, i.e., $m=5.0$, $6.0$, and $7.0$. The
distributions $P(S)$ and $n(s)$ are obtained from the dynamics
up to $t=5\times 10^{6}$. In both panels, the distributions
for $m=6.0$ and $7.0$ are vertically shifted for clarity.}
\label{fig5}
\end{figure}
The distributions of the avalanche size $S$ for several
values of the node tolerance parameter $m$ are presented in
Fig.~\ref{fig5}(a). The avalanche size distribution $P(S)$
also follows a power-law relation, i.e.,
\begin{equation}
P(S)\propto S^{-\lambda}.
\label{eq:P(S)}
\end{equation}
Our result indicates that the exponent $\lambda$ does not depend
on $m$ and is estimated as $\lambda=2.60\pm 0.02$ for $m=7.0$.
The rightmost hump in $P(S)$ at around $S\sim 900$ represents
the contribution from complete collapses. The second hump from
the right corresponds to decays of large networks to assemblies
of dimers. Moreover, the broad hump near $S=150$ found in the
result for $m=5.0$ is related to the hump in $W(T)$ shown by
Fig.~\ref{fig4}. This broad hump represents the typical
avalanche size of cascades from typical networks with
$N_{\text{typ}}\sim 700$ nodes. Therefore, these humps are
attributable to finite-size effects associated with our choice
of the model parameters.

Figure \ref{fig5}(b) shows the cluster size distribution $n(s)$
for three different values of $m$. The cluster size $s$ at time
$t$ is the number of nodes in a component included in the
network $\mathcal{G}(t)$. The cluster size distribution function
$n(s)$ is calculated from all components of the network at every
time step in the entire dynamics. The distribution $n(s)$ has a
power-law form,
\begin{equation}
n(s)\propto s^{-\tau},
\label{eq:n(s)}
\end{equation}
as well as $W(T)$ and $P(S)$. The exponent $\tau$, calculated
as $\tau=2.92\pm 0.02$ for $m=7.0$, is also independent of $m$.
In contrast to the distribution $P(S)$, the influence of complete
collapses to $n(s)$ is inconspicuous. This is because the size of
a network just after a complete collapse is $N_{\text{ini}}(=3)$
and the number of components with $s=3$ generated in the whole
period of the network evolution is extremely large compared to
the number of complete collapses occurring in the same period.

All the above results, namely the intermittency of $S(t)$ and
the power-law forms of $W(T)$, $P(S)$, and $n(s)$, strongly
support that the dynamics of network structure in our model
exhibits SOC behavior. These results also show that the
universality class of self-organized criticality does not
depend on the node tolerance parameter $m$. The relation to
other parameters will be discussed later.

\subsection{Fractal and small-world networks}
\label{subsec:Fractal}

As we mentioned in Sect.~\ref{subsec:instability}, a cascade of
overload failures gives a fatal damage to a connected network
$\mathcal{G}_{0}$ of size $N_{0}$ if the load reduction
parameter $r$ is less than $r_{\text{c}}(N_{0})$, and the giant
component after a cascade at $r=r_{\text{c}}(N_{0})$ has a
fractal structure \cite{Mizutaka15}. In the present SOC model,
on the other hand, $r$ decreases with the network size
$N_{0}(t)$. For $r[N_{0}(t)]$ chosen as Eq.~(\ref{r(N)}), the
parameter $r$ decreases from a large enough value
$r_{\text{max}}$ for $N_{0}(t)\le N_{\text{ini}}$ to zero for
$N_{0}(t)\ge N_{\text{max}}$. Since any network is completely
collapsed by a cascade of failures at $r=0$, through the
network growth, $r[N_{0}(t)]$ must eventually encounter the
\textit{critical} value $r_{\text{c}}$ at which the cascade of
failures provides the percolation transition of the network.
(Precisely speaking, the term ``critical" is not appropriate
because $N_{0}(t)$ is finite. However, we use this terminology
by supposing the case that sufficiently large networks are
generated under suitable values of the model parameters. The
word ``critical" in the rest of this paper will be used in the
same sense.) If a network with the size $N_{0}(t)$ satisfying
$r[N_{0}(t)]=r_{\text{c}}$ experiences a cascade of overload
failures, we expect that the giant component after the cascade
has a fractal structure.

In the case of cascading failures starting with a fixed
connected network $\mathcal{G}_{0}$ of size $N_{0}$ in which
all the capacities of nodes are definitely determined by their
degrees and the initial total load $W_{0}$, the value of
$r_{\text{c}}(N_{0})$ is theoretically calculated as addressed
in the last paragraph of Sect.~\ref{subsec:instability} \cite{Mizutaka15}.
In our SOC model, however, the capacity of a node depends on
the total load at the time when the node was introduced in the
system. Thus, the node capacities in the network
$\mathcal{G}_{0}(t)$ depends strongly on the past history of
$\mathcal{G}_{0}(t)$, and the critical load reduction parameter
$r_{\text{c}}$ cannot be uniquely determined by the size of
$\mathcal{G}_{0}(t)$. Since the theoretical method proposed by
Ref.~\ref{mizu} is not applicable to dynamics governed by such
hysteresis effects, we need to examine numerically whether
$r[N_{0}(t)]$ of the network $\mathcal{G}_{0}(t)$ is close to
an unknown value of $r_{\text{c}}$ peculiar to
$\mathcal{G}_{0}(t)$.

\begin{figure}[ttt]
\begin{center}
\includegraphics[width=0.4\textwidth]{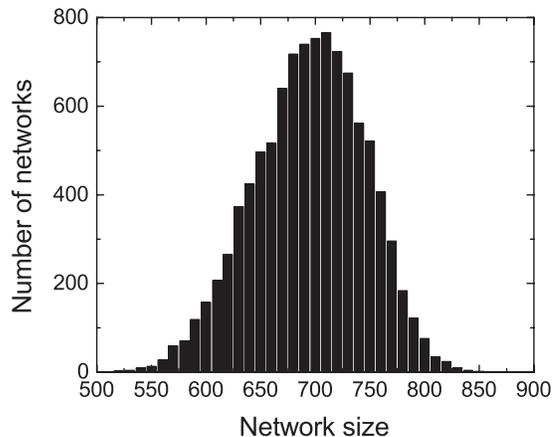}
\end{center}
\caption{Histogram of the number of pre-critical networks just
before critical cascades versus the size of the pre-critical
network. The histogram is obtained for the first $10^{4}$
critical cascades occurring in the dynamics under the condition
$m=5.0$.}
\label{fig6}
\end{figure}
If $r[N_{0}(t)]$ of the network $\mathcal{G}_{0}(t)$ is much
larger than $r_{\text{c}}$, a cascade of overload failures, if
any, eliminates only a small fraction of nodes from
$\mathcal{G}_{0}(t)$ and does not change the giant component
size so much. On the other hand, a cascade with
$r[N_{0}(t)]\ll r_{\text{c}}$ causes a complete collapse of
$\mathcal{G}_{0}(t)$. If $r[N_{0}(t)]$ is close to $r_{\text{c}}$,
the cascade is marginal, for which the size of the giant
component after the cascade must be much smaller than the
original giant component size of $\mathcal{G}_{0}(t)$ but
still much larger than $N_{\text{ini}}$. From the above
consideration, we regard in this work a cascade of overload
failures at time $t$ satisfying the following conditions as a
\textit{critical cascade} whose load reduction parameter should
be close to $r_{\text{c}}$:
\begin{equation}
\displaystyle
\frac{N_{\text{LC}}(t)}{N_{\text{LC}}(t-1)}\le 0.5
\quad \text{and}\quad
N_{\text{LC}}(t)\ge 100 ,
\label{eq:cr_cascade}
\end{equation}
where $N_{\text{LC}}(t)$ is the number of nodes in the largest
component of $\mathcal{G}(t)$. The specific values $0.5$ and
$100$ in Eq.~(\ref{eq:cr_cascade}) are not important as long as
$N_{\text{LC}}(t)/N_{\text{LC}}(t-1)$ and $N_{\text{LC}}(t)$
are much smaller and larger than $1$, respectively. In the
sense of the percolation transition by cascading overload
failures, a network after completing a critical cascade can be
considered as a \textit{critical network} $\mathcal{G}_{\text{c}}$.
Also, we call a network just before a critical cascade a
\textit{pre-critical network} $\mathcal{G}_{\text{pre}}$.
Figure \ref{fig6} shows the histogram of the number of
pre-critical networks as a function of the size of
$\mathcal{G}_{\text{pre}}$. This result indicates that critical
cascades are likely to occur on networks of size $N(t)\sim 700$
for the present parameter set. Considering that the most
probable network size $N_{\text{typ}}$ is also about $700$ as
shown by the inset of Fig.~\ref{fig4}, critical cascades take
place frequently during SOC dynamics. This means that critical
networks are generated very often by such cascades.

\begin{figure}[ttt]
\begin{center}
\includegraphics[width=0.48\textwidth]{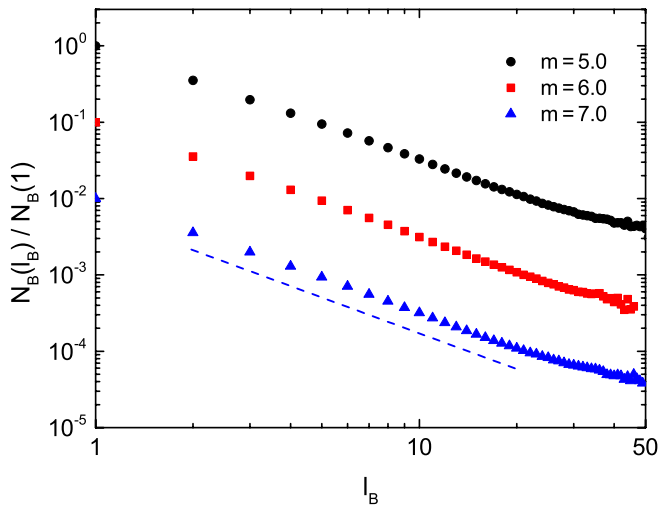}
\end{center}
\caption{(Color online) $N_{\text{B}}(l_{\text{B}})$ for giant
components in critical networks generated by SOC dynamics under
the conditions $m=5.0$, $6.0$, and $7.0$. The longitudinal axis
indicates $N_{\text{B}}(l_{\text{B}})/N_{\text{B}}(1)$ averaged
over $1,000$ realizations of critical networks. The results for
$m=6.0$ and $7.0$ are vertically shifted for clarity. The
straight dashed line has the slope $d_{\text{B}}=1.53$ which is
obtained by the least-squares fit for the data for $m=7.0$ from
$l_{\text{B}}=2$ to $20$.}
\label{fig7}
\end{figure}
We study the fractal property of giant components in critical
networks. As we explained in Sect.~\ref{sec:intro}, if a given
connected network is fractal, the minimum number
$N_{\text{B}}(l_{\text{B}})$ of subgraphs of diameter less than
$l_{\text{B}}$ required to cover the network satisfies the
relation
\begin{equation}
N_{\text{B}}(l_{\text{B}})\propto l_{\text{B}}^{-d_{\text{B}}},
\label{eq:fractal}
\end{equation}
where $d_{\text{B}}$ is the fractal dimension of the network
\cite{Song05}. We calculate $N_{\text{B}}(l_{\text{B}})$ for
giant components in $1,000$ critical networks appearing in the
dynamics by using the compact-box-burning algorithm \cite{Song07}
and average $N_{\text{B}}(l_{\text{B}})/N_{\text{B}}(1)$ over
these realizations, where $N_{\text{B}}(1)$ is equal to the
number of nodes in the largest component. The results for
$m=5.0$, $6.0$, and $7.0$ are plotted in Fig.~\ref{fig7}. These
plots clearly demonstrate that the quantity
$N_{\text{B}}(l_{\text{B}})$ satisfies Eq.~(\ref{eq:fractal})
and the fractal dimension $d_{\text{B}}$ does not depend on
$m$. The value of $d_{\text{B}}$ estimated from the result for
$m=7.0$ is $1.53\pm 0.01$. It is interesting that this fractal
dimension is close to $d_{\text{B}}=1.54\pm 0.01$ that has been
computed for the giant component after a critical cascade
starting with an Erd\H{o}s-R\'enyi (ER) random graph \cite{Mizutaka15}.
The topology of a pre-critical network $\mathcal{G}_{\text{pre}}$
in SOC dynamics is not the same as that of the ER random graph
$\mathcal{G}_{\text{ER}}$. In addition, the capacity of a node
in $\mathcal{G}_{\text{pre}}$ depends on the total load of
the system when the node was introduced, while the node capacity
in $\mathcal{G}_{\text{ER}}$ is determined by the degree of the
node and a fixed initial total load $W_{0}$. In spite of these
discrepancies, it is not surprising that both fractal
dimensions are the same, which implies the same universality
class between percolation transitions for
$\mathcal{G}_{\text{pre}}$ and $\mathcal{G}_{\text{ER}}$. This
is because the degree distribution of $\mathcal{G}_{\text{pre}}$
has an exponential tail, like that of $\mathcal{G}_{\text{ER}}$,
and the node capacity distribution is not wide, which behaves
similarly to the distribution of $N(t)$ shown in the inset of
Fig.~\ref{fig4}.

\begin{figure}[ttt]
\begin{center}
\includegraphics[width=0.48\textwidth]{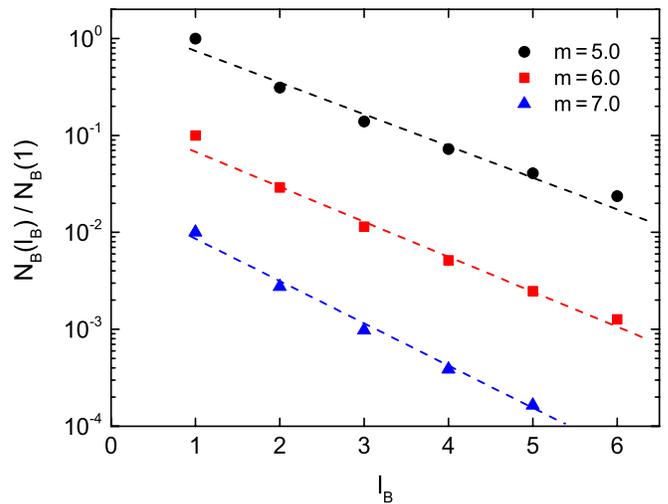}
\end{center}
\caption{(Color online) $N_{\text{B}}(l_{\text{B}})$ for largest
components in networks that first reach the size $N(t)=300$
after complete collapses in SOC dynamics under the conditions
$m=5.0$, $6.0$, and $7.0$. The longitudinal axis indicates
$N_{\text{B}}(l_{\text{B}})/N_{\text{B}}(1)$ averaged over $1,000$
realizations of such networks. The results for $m=6.0$ and $7.0$
are vertically shifted for clarity. The dashed lines are guides
to the eye.}
\label{fig8}
\end{figure}
Let us examine $N_{\text{B}}(l_{\text{B}})$ for off-critical
networks. If a network is far from criticality, we can expect
that the network has a small-world structure, because the
network formed by random attachment of new nodes has many
short-cut edges. For a small-world network, the number of
covering subgraphs $N_{\text{B}}(l_{\text{B}})$ decreases
exponentially with $l_{\text{B}}$, namely,
\begin{equation}
N_{\text{B}}(l_{\text{B}})\propto \exp(-l_{\text{B}}/l_{0}),
\label{eq:small_world}
\end{equation}
where $l_{0}$ is a characteristic path length. We calculated
$N_{\text{B}}(l_{\text{B}})$ for networks (or their largest
components if not connected) that first reach the size
$N(t)=300$ after complete collapses, and averaged
$N_{\text{B}}(l_{\text{B}})/N_{\text{B}}(1)$ over $1,000$
realizations of such networks in SOC dynamics. The results
shown in Fig.~\ref{fig8} indicate the small-world property of
these networks. Our SOC model thus generates both fractal
and small-world networks in a \textit{single} dynamics.

\begin{figure}[ttt]
\begin{center}
\includegraphics[width=0.48\textwidth]{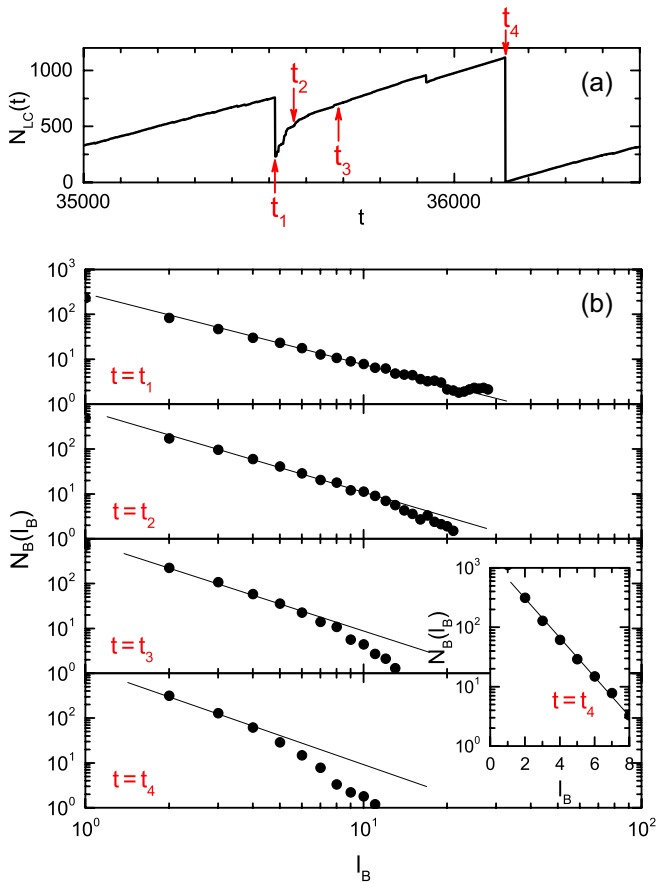}
\end{center}
\caption{(Color online) (a) Time dependence of the number of
nodes $N_{\text{LC}}(t)$ in the largest component of the network
$\mathcal{G}(t)$ in the dynamics for $m=7.0$. The arrows
indicate the times at which $N_{\text{B}}(l_{\text{B}})$'s are
calculated. (b) $N_{\text{B}}(l_{\text{B}})$'s for largest
components in $\mathcal{G}(t_{1})$, $\mathcal{G}(t_{2})$,
$\mathcal{G}(t_{3})$, and $\mathcal{G}(t_{4})$ from top to
bottom. Thin lines are guides to the eye. The rapid decrease of
$N_{\text{B}}(l_{\text{B}})$ for $l_{\text{B}}\gg
l_{\text{co}}$ at $t=t_{3}$ or $t_{4}$ indicates that the
largest component has a small-world structure in a longer
length-scale than $l_{\text{co}}$. The inset shows
$N_{\text{B}}(l_{\text{B}})$ at $t=t_{4}$ in a semi-logarithmic
scale.}
\label{fig9}
\end{figure}
We further investigate the crossover behavior from fractal to
small-world structure associated with the time evolution from a
critical network. Figure \ref{fig9} illustrates a typical
profile change of $N_{\text{B}}(l_{\text{B}})$ for networks
formed at several times from $t_{1}$ at which a critical
network appears to $t_{4}$ just before the next complete
collapse. The times at which $N_{\text{B}}(l_{\text{B}})$'s are
calculated are indicated in Fig.~\ref{fig9}(a) by arrows on the
time dependence of the largest component size
$N_{\text{LC}}(t)$. At $t=t_{1}$, $N_{\text{B}}(l_{\text{B}})$
follows a power law, which suggests that the giant component in
$\mathcal{G}(t_{1})$ has a fractal structure as we expect.
After this time, the largest component size rapidly increases
as shown in Fig.~\ref{fig9}(a). This is because newly added
nodes are more likely to merge separated fractal components but
less likely to be connected onto a single component. Therefore,
the largest component at $t=t_{2}$ remains fractal at almost
any scale. When the time elapses further, the increase of
$N_{\text{LC}}(t)$ becomes moderate. This implies that the
merging process of separated components has been mostly
finished and new nodes are simply incorporated in the largest
component. In this case, newly added nodes bring short-cut
edges in the largest component, which makes the network
small-world as shown by $N_{\text{B}}(l_{\text{B}})$ at
$t=t_{3}$ in Fig.~\ref{fig9}(b). More precisely, the network is
small-world in a longer length scale than the average distance
$l_{\text{co}}$ between terminal nodes of short-cut edges
introduced by new nodes, while it is fractal for
$l_{\text{B}}\ll l_{\text{co}}$. This situation is similar to
the case that a lattice-like network changes into a small-world
one by random rewirings in the Watts-Strogatz model
\cite{Barthelemy99} In fact, a high density of short-cut edges
at $t=t_{4}$ reduces the crossover length $l_{\text{co}}$ and
the small-world property can be found in the whole $l_{\text{B}}$
range as shown by the inset of Fig.~\ref{fig9}(b).

\subsection{Suitable choice of parameter values}
\label{subsec:Parameters}

All the above arguments are based on specific values of the
model parameters. If the dynamical properties presented above
are peculiar to these parameter values, it cannot be said that
the present model exhibits self-organized criticality, because
of the necessity of tuning the external parameters. It has,
however, been confirmed that the results are essentially
independent of the choice of parameter values if these
parameters lie in \textit{suitable ranges}. In this subsection,
we discuss the suitable parameter ranges to realize SOC dynamics.

To find the suitable ranges of parameter values, let us
consider how large a network could grow if the system did not
experience any critical cascades and complete collapses. Even
in this case, a network cannot grow infinitely. The expectation
number $\langle S\rangle$ of eliminated nodes per unit time
step increases with the network size $N(t)$, and eventually
$\langle S\rangle$ reaches the incrementation of $N(t)$ at
every time step due to the participation of a new node. Once
this is the case, the network does not grow any more. The
network size $N(t)$ then fluctuates around a stationary size
with small amplitudes. The stationary size $N_{\text{st}}$ can
be roughly estimated by the overload probability. In the
absence of critical cascades and complete collapses, we can
consider approximately that all cascading overload failures
stop at the first step of the cascade process and subsequent
avalanches triggered by the first failures do not occur,
because avalanche sizes are small. This approximation enables
us to calculate the steady-state expectation number of failed
nodes per unit time step by
\begin{equation}
\langle S\rangle = N_{\text{st}}\sum_{k}\mathcal{P}_{\text{st}}(k)
F_{W(N_{\text{st}})}(k),
\label{Sav}
\end{equation}
where $\mathcal{P}_{\text{st}}(k)$ is the degree distribution
of a steady-state network $\mathcal{G}_{\text{st}}$ and
$F_{W(N_{\text{st}})}(k)$ is the overload probability of a node
of degree $k$ in $\mathcal{G}_{\text{st}}$. With the aid of the
regularized incomplete beta function, the probability
$F_{W(N_{\text{st}})}(k)$ is, with reference to
Eq.~(\ref{eq:FWk}), given by
\begin{equation}
F_{W(N_{\text{st}})}(k)=I_{k/\mu N_{\text{st}}}
\left[\lfloor q_{\mu}(N_{\text{st}})\rfloor +1,W(N_{\text{st}})-
\lfloor q_{\mu}(N_{\text{st}})\rfloor \right],
\label{eq:Fst}
\end{equation}
where
\begin{equation}
q_{\mu}(N_{\text{st}})=
\frac{a\mu}{2}+m\sqrt{\frac{a\mu}{2}\left(1-\frac{1}{N_{\text{st}}}\right)},
\label{eq:qst}
\end{equation}
and
\begin{equation}
W(N_{\text{st}})=\frac{a\mu N_{\text{st}}}{2}.
\label{eq:Wst}
\end{equation}
Here, we approximated the average degree of
$\mathcal{G}_{\text{st}}$ by $\mu$ for the reason that nodes
with degree greater than $\mu$ are more likely to be eliminated
by overload failures in $\mathcal{G}_{\text{st}}$ while the
average degree would be larger than $\mu$ (equal to $2\mu$) if
the network monotonically grew without any node elimination. In
the steady state, $\langle S\rangle$ must be equal to the
incrementation of the network size per unit time step, namely
$1$. Therefore, the stationary size $N_{\text{st}}$ is
determined by the relation,
\begin{equation}
\displaystyle
N_{\text{st}}=\frac{1}{\sum_{k}\mathcal{P}_{\text{st}}(k)
F_{W(N_{\text{st}})}(k)}.
\label{stationary_size}
\end{equation}
If we neglect critical cascades and complete collapses, the
network can grow up to the size $N_{\text{st}}$ obtained by
solving the above transcendental equation. But actually,
the network encounters critical cascades or complete
collapses before reaching this size if $N_{\text{st}}$ is
larger than the typical size $N_{\text{pre}}$ of pre-critical
networks $\mathcal{G}_{\text{pre}}$. In this case, and only
in this case, critical cascades generate fractal networks,
and the present model exhibits SOC character. Otherwise,
critical cascades themselves never take place in the dynamics.
Thus, the condition to realize SOC dynamics is
\begin{equation}
N_{\text{st}} \gg N_{\text{pre}} \gg 1,
\label{condition}
\end{equation}
where the condition $N_{\text{pre}} \gg 1$ guarantees that the
system is large enough to exhibit genuine self-organized
criticality. What is the relation between the above condition
and the model parameters? Among several parameters
characterizing our model, parameters related to the initial
network $\mathcal{G}(0)$, namely, $N_{\text{ini}}$,
$M_{\text{ini}}$, and the topology of $\mathcal{G}(0)$, are
obviously irrelevant to the condition (\ref{condition}). It is
thus significant to elucidate how $N_{\text{st}}$ and
$N_{\text{pre}}$ depend on $a$ (the load carried by a single
edge), $m$ (the node tolerance parameter), $\mu$ (the number of
edges of a newly added node), and the functional form of the
load reduction parameter $r(N)$.

The stationary size $N_{\text{st}}$ depends on $a$, $m$, and
$\mu$. Equation (\ref{stationary_size}) reveals the relation of
$N_{\text{st}}$ to these parameters. Since the preferential
elimination of nodes with degree much larger than $\mu$ in
$\mathcal{G}_{\text{st}}$ gives a sharp peak of
$\mathcal{P}_{\text{st}}(k)$ at $k=\mu$, $\mathcal{P}_{\text{st}}(k)$
hardly depends on $N_{\text{st}}$. Furthermore, the overload
probability $F_{W(N_{\text{st}})}(k)$ presented by
Eq.~(\ref{eq:Fst}) indeed depends only very weakly on
$N_{\text{st}}$, which comes from the property of the
regularized incomplete beta function. Therefore,
Eq.~(\ref{stationary_size}) is not actually transcendental, and
$N_{\text{st}}$ can be evaluated by $\mathcal{P}_{\text{st}}(k)$
and $F_{W(N)}(k)$ for a haphazardly chosen value of $N(\gg
\mu)$. The probability $F_{W(N)}(k)$ given by
Eqs.~(\ref{eq:Fst})-(\ref{eq:Wst}) with $N_{\text{st}}$
replaced by $N$ is a decreasing function of $m$ and $\mu$ for
any $k$. Hence, $N_{\text{st}}$ obtained by
Eq.~(\ref{stationary_size}) increases with $m$ and $\mu$
regardless of $\mathcal{P}_{\text{st}}(k)$. The $a$ dependence
of $F_{W(N)}(k)$ is, however, influenced by the value of $k$.
$F_{W(N)}(k)$ increases with $a$ if $k>\mu$, while it decreases
for $k<\mu$. Meanwhile, for a fixed value of $a$, $F_{W(N)}(k)$
for $k<\mu$ is negligibly small. Thus, $F_{W(N)}(k)$ for $k$
larger than $\mu$ dominates the summation in
Eq.~(\ref{stationary_size}), independently of the form of
$\mathcal{P}_{\text{st}}(k)$. This fact and the property of
$F_{W(N)}(k)$ of being an increasing function of $a$ for
$k>\mu$ show that $N_{\text{st}}$ decreases with $a$.
Consequently, we need to choose large values of $m$ and $\mu$
and a small value of $a$ to obtain large $N_{\text{st}}$.

On the other hand, $N_{\text{pre}}$ is the typical size of a
network whose load reduction parameter $r$ is equal to the
critical value $r_{\text{c}}$ specific to the network. The
parameter $r$ is uniquely determined by the network size $N$,
while $r_{\text{c}}$ depends not only on $N$ but also on the
past history of the network. Approximating the typical size of
pre-critical networks by the size of a typical pre-critical
network $\mathcal{G}_{\text{pre}}$, $N_{\text{pre}}$ must
satisfy
\begin{equation}
r(N_{\text{pre}})=r_{\text{c}}[\mathcal{G}_{\text{pre}}(N_{\text{pre}})],
\label{eq:r=rpre}
\end{equation}
where $r_{\text{c}}[\mathcal{G}_{\text{pre}}(N_{\text{pre}})]$
is the critical load reduction parameter of
$\mathcal{G}_{\text{pre}}$ whose size is $N_{\text{pre}}$.
Since the right-hand size of Eq.~(\ref{eq:r=rpre}) is a
function of $a$, $m$, $\mu$, and $N_{\text{pre}}$, the size
$N_{\text{pre}}$ as the solution of Eq.~(\ref{eq:r=rpre})
depends on these parameters in addition to the functional form
of $r(N)$. If $r(N)$ decreases slowly with $N$, however, the
solution $N_{\text{pre}}$ is mainly governed by the form of $r(N)$
rather than the precise value of $r_{\text{c}}$. In order to
satisfy $N_{\text{pre}} \gg 1$, $r(N)$ needs to decrease very
slowly with the network size. In the case that $r(N)$ is set as
Eq.~(\ref{r(N)}), $N_{\text{max}}$ must be chosen to be large
enough.

In conclusion, the load reduction parameter $r(N)$ must
decrease with $N$ very slowly to realize the condition
$N_{\text{pre}} \gg 1$, and the node tolerance parameter $m$
and the load by edge $a$ should be large and small enough,
respectively, so that $N_{\text{st}}$ becomes much larger than
$N_{\text{pre}}$. Although a large value of $\mu$ is preferable
for the condition $N_{\text{st}}\gg N_{\text{pre}}$, results
are not strongly influenced by $\mu$ because the number of
edges of a new node is always restricted by $2\le \mu \le
N_{\text{ini}}$ with small $N_{\text{ini}}$. Our choice of
values for $a$, $m$, and $\mu$ in this section obviously
satisfies the condition $N_{\text{st}} \gg N_{\text{pre}}$,
because we have critical cascades in the dynamics. In fact,
$N_{\text{st}}$ estimated by Eq.~(\ref{stationary_size}) with
the numerically obtained $\mathcal{P}_{\text{st}}(k)$ is $861$
for $a=2.0$, $\mu=2$, and $m=5.0$, which is larger than
$N_{\text{pre}}\simeq 700$ as indicated in Fig~\ref{fig6}. We
have confirmed that the universality class of self-organized
criticality, namely the set of the exponents $\eta$, $\lambda$,
$\tau$, and $d_{\text{B}}$, does not depend on the choice of
parameter values if the condition (\ref{condition}) is
satisfied. It has also been checked that the functional form of
$r(N)$ is irrelevant to SOC dynamics as far as $r(N)$ is a
slowly decreasing function of $N$.

\section{Summary}
\label{sec:summary}

We have proposed a model of self-organized critical (SOC)
dynamics of complex networks and presented a possible
explanation of the emergence of fractal and small-world
networks. Our model combines a network growth and its decay due
to the instability of large grown networks against cascading
overload failures. Cascading failures occur intermittently and
prevent networks from growing infinitely. The distribution of
the inactive time interval between successive cascades of
failures has a power-law form. Both the avalanche size that is
the number of eliminated nodes in a single cascade and the
cluster size defined as the number of nodes in a connected
component also obey power-law distributions. These facts indicate
that the network dynamics possesses SOC characteristics. During
the SOC dynamics, the load reduction parameter $r$ varies with
the network size. When $r$ of the network coincides with its
critical value $r_{\text{c}}$, a cascade of overload failures
(critical cascade) decays the network into a critical one. We
have shown that giant components just after critical cascades
have fractal structures. The fractal dimension $d_{\text{B}}$
is close to that for the giant component after a critical
cascade starting with an Erd\H{o}s-R\'enyi random graph. In
contrast, networks far from criticality display the small-world
property. In particular, we demonstrated the crossover behavior
from fractal to small-world structure in a growing process from
a critical network, which is caused by short-cut edges
introduced by newly added nodes. We have also discussed suitable
parameter values to realize SOC dynamics.

It is significant to notice that the present model is somewhat
different from previous SOC models. In a conventional SOC
model, a routine procedure in the dynamics, such as placement
of grains of sand in the sandpile model \cite{Bak87} or renewals
of fitness values in the Bak-Sneppen model \cite{Bak93}, takes a
system close to the critical point, but the instability of
critical or near-critical states drives the system away from
criticality accompanied by some sort of avalanches. In our
model, on the other hand, the network growth as a routine
procedure takes the system away from the critical point, but
the instability of large grown networks makes the network
critical. Although the roles of growth and instability are
opposite to those of conventional models, the system described
by our model exhibits the most of SOC characteristics as
explained in Sect.~\ref{sec:results}. This implies that the
present model provides a new type of self-organized
criticality. Our model generates non-scale-free networks with
homogeneous degree distributions and belongs to a specific
universality class of SOC dynamics, independently of the choice
of values of the model parameters. It is then interesting to
study how the model should be modified to belong to another
class of self-organized criticality with forming scale-free
networks.

\begin{acknowledgments}
This work was supported by a Grant-in-Aid for Scientific
Research (Nos.~25390113 and 14J01323) from the Japan Society
for the Promotion of Science. Numerical calculations in this
work were performed in part on the facilities of the
Supercomputer Center, Institute for Solid State Physics,
University of Tokyo.
\end{acknowledgments}

\end{document}